\documentclass[prl,twocolumn,showpacs,
superscriptaddress,floatfix]{revtex4}
\usepackage{graphicx}

\begin{document}

\title{Statistics of the Mesoscopic Field}

\author{A.A.~Chabanov}
\altaffiliation{Electronic address: a-chabanov@northwestern.edu}
\affiliation{Department of Physics, Queens College of the City
University of New York, Flushing, NY 11367, USA}
\affiliation{CEMS, University of Minnesota, Minneapolis, MN 55455, USA}

\author{A.Z.~Genack}
\affiliation{Department of Physics, Queens College of the City
University of New York, Flushing, NY 11367, USA}

\date{14 February, 2005}

\begin{abstract}
We find in measurements of microwave transmission through quasi-1D 
dielectric samples for both diffusive and localized waves that the 
field normalized by the square root of the spatially averaged flux 
in a given sample configuration is a Gaussian random process with 
position, polarization, frequency, and time. As a result, the probability 
distribution of the field in the random ensemble is a mixture of 
Gaussian functions weighted by the distribution of total transmission, 
while its correlation function is a product of correlators of the 
Gaussian field and the square root of the total transmission.
\end{abstract}

\maketitle
Enhanced fluctuations and correlation of the local and 
spatially averaged electronic and electromagnetic flux in 
mesoscopic samples have been extensively studied in the 
last two decades \cite{Mesobook,RMP,Bartbook}. Yet 
the impact of weak and strong localization upon the statistics 
of the field itself has not been investigated. The field at a 
point in a multiple-scattering medium is the superposition of 
partial waves associated with all wave trajectories or Feynman 
paths reaching that point. The changing distribution of paths 
with position produces a speckled intensity pattern which 
fluctuates on a scale of the wavelength, $\lambda$. When a 
large number of statistically independent terms contribute 
to the field and the sample is far from the localization 
threshold so that long-range correlation is not substantial, 
the distribution of in- and out-of-phase components of the 
field is Gaussian \cite{Goodman}. This leads to a negative 
exponential distribution for the intensity of a polarized 
component of the field, which is widely observed in diffusive 
media. 

In mesoscopic samples, however, the crossing of Feynman paths 
gives rise to correlation in the intensity at distant points. 
This leads to enhanced fluctuations of intensity 
\cite{Rossum95,Kogan95,Beenakker96,Marin99}, total transmission 
\cite{corr,Polkosnik90,Marin97}, and conductance \cite{UCF,C3}, 
with increasing variances as the localization threshold is 
approached. Since the statistics of transport in mesoscopic 
samples reflects the statistics of the underlying field, it 
is of interest to investigate field statistics in samples 
with increasing intensity correlation. 

In this Letter, we investigate the structure of the 
probability distribution and correlation of the mesoscopic 
field in both the frequency and time domain for diffusive and 
localized waves. We find that the distribution of the field 
in a given realization of a random quasi-1D sample and in the 
subset of realizations with the same value of total 
transmission is Gaussian, even for strongly localized waves. The 
field distribution in the random ensemble is therefore a mixture of 
Gaussian functions weighted by the distribution of total 
transmission. Thus the statistics of the total transmission 
can be obtained from the statistics of the field, as well 
as the reverse. When the field is normalized by the square 
root of the spatially averaged flux in each configuration, 
it becomes a Gaussian random process with position, polarization, 
frequency, and time. The statistical independence of the normalized 
Gaussian field and square root of the total transmission 
allows the field correlation function to be written as a 
product of their respective correlation functions. The field 
correlation function with frequency or time shift exhibits 
the impact of localization through the corresponding correlation 
function of the square root of the total transmission, 
while the field correlation function with displacement and
polarization rotation has a simple universal form in both the 
frequency and time domain for diffusive and localized waves.

To examine field statistics, we measure microwave spectra 
of the field transmission coefficient in ensembles of random 
quasi-1D samples of alumina spheres with the use of a vector 
network analyzer. Alumina spheres with diameter 0.95 cm and 
refractive index 3.14 are embedded in Styrofoam spheres of 
refractive index 1.04 to produce an alumina volume fraction 
of 0.068. The spheres are contained in a copper tube of 
diameter 7.3 cm and length 61 cm with reflecting sidewalls 
and open ends. Spectra are taken in $10^4$ sample configurations 
produced by rotating the sample tube, in steps of 0.3 MHz in 
the frequency intervals 14.7-15.7 GHz (measurement A) and 
9.95-10.15 GHz (measurement B), in which waves are diffusive 
and localized, respectively \cite{Chabanov01}. The frequency 
intervals are sufficiently narrow that propagation parameters 
change little within each interval. 

The probability distributions of the transmitted intensity 
and total transmission normalized by their ensemble averages, 
$s_{ab}$=$|t_{ab}|^2/\langle|t_{ab}|^2\rangle$ and 
$s_{a}$=$\Sigma_b|t_{ab}|^2/\langle\Sigma_b|t_{ab}|^2\rangle$, 
respectively, where $t_{ab}$ is the transmission coefficient 
for incident field $a$ and outgoing field $b$, are determined 
by the variance of the normalized total transmission, var$(s_a)$ 
\cite{Marin97,Marin99}. Var$(s_{ab})$ can be found from the 
relation, var$(s_{ab})$=1+2var$(s_a)$ \cite{Kogan95}. In the 
limit of Gaussian field statistics for the random ensemble, 
var$(s_{ab})$=1 \cite{Goodman}. In measurements A and B, 
var$(s_{ab})$=1.18 and 6.18, respectively. In order to study 
field statistics in even more strongly correlated samples, 
we examine the statistics of pulsed transmission of localized 
waves for a long delay from an exciting pulse, since the 
degree of correlation increases with time delay \cite{Polar04,Skip04}. 
The response to a pulse with a Gaussian temporal envelope of 
width $\sigma_{t}$=160 ns is obtained from the Fourier transform 
of the product of the field transmission spectra of measurement 
B and a Gaussian spectral function of bandwidth 
$\sigma$=$1\textrm{MHz}\approx0.6\delta\nu$, where $\delta\nu$ 
is the field correlation frequency \cite{Drake}. The field 
statistics are examined for waves delayed by 740 ns from the 
center of exciting pulse. In this case (measurement C), 
var$[s_{ab}(t)]$=20.1. 

We first consider the probability distribution of the field 
transmission coefficient normalized to the square root of 
the ensemble-averaged intensity, 
$E$=$t_{ab}/\!\sqrt{\langle|t_{ab}|^2\rangle}$. We find in 
all cases that the statistics of the real and imaginary parts 
of the field, $r$=Re[$E$] and $i$=Im[$E$], respectively, are 
the same and that $r$ and $i$ have zero mean and vanishing 
cross correlation. The distributions $P(\alpha)$, where 
$\alpha$=$r,i$, are shown by the solid curves in Fig.~1. 
Increasing deviations from a Gaussian distribution, 
$P(\alpha)$=${1\over\sqrt{\pi}}\exp(-\alpha^2)$, 
obtained for the model of the field as a random phasor 
sum (dashed curve) \cite{Goodman}, are seen for the 
distributions in measurements with larger values of 
var$(s_{ab})$.  
\begin{figure}[b!]
\includegraphics[width=\columnwidth]{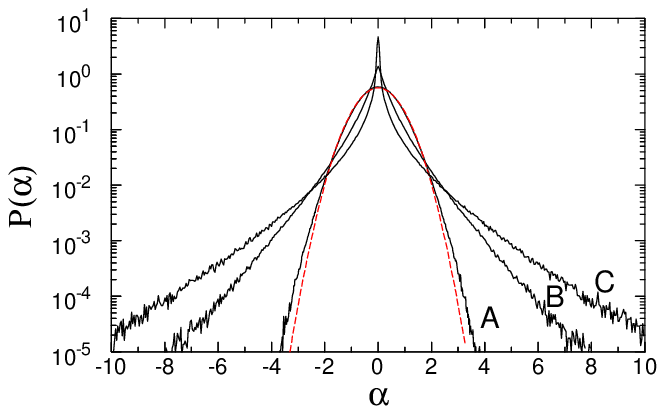}
\caption{Field distributions $P(\alpha)$ of measurements A, B, 
and C (solid). The dashed curve is the Gaussian distribution, 
$P(\alpha)$=${1\over\sqrt{\pi}}\exp(-\alpha^2)$, of the model 
of the field as a random phasor sum \cite{Goodman}.}
\end{figure}

Though departures from a Gaussian distribution in ensembles 
of mesoscopic samples arise from extended correlation generated 
by the crossing of wave trajectories, we expect that the field 
distribution in a given quasi-1D sample configuration is Gaussian. 
Once wave trajectories cross in the medium, intensity fluctuations 
propagate throughout the sample, leading to correlation in the 
field amplitude on the output surface at points separated by 
considerably more than the field correlation length of $\lambda/2$ 
\cite{Wolf,Shapiro}. This gives rise to enhanced fluctuations of 
the total transmission from configuration to configuration. 
However, in a given configuration the field at each point on 
the output surface is the sum of a large number of partial 
waves with identically distributed amplitudes and phases, which is
due to perfect mixing of modes in a quasi-1D sample. When 
the number of field coherence areas on the output surface 
is large, the field in a given configuration and in the 
subset of configurations with a specific value of $s_a$ 
is a Gaussian random variable. We therefore expect that 
the joint distribution of the real and imaginary parts of 
the field in the random ensemble, $P(r,i)$, will be a mixture 
of Gaussian distributions \cite{Mittleman} of zero means and 
variances $s_{a}/2$, with mixing proportions given by the 
total transmission distribution $P(s_{a})$,
\begin{equation}
P(r,i)=\!\int_{0}^{\infty}\!\!ds_{a}P(s_{a}){1\over
\pi s_{a}}\exp\!\left(-{r^2+i^2\over s_{a}}\right).
\end{equation}

Since $s_{ab}=|E|^2\equiv r^{2}+i^{2}$, the distribution 
of normalized intensity, $P(s_{ab})$, is a mixture of 
negative exponential distributions,
\begin{equation}
P(s_{ab})=\!\int_{0}^{\infty}\!\!ds_{a} P(s_{a})
{1\over s_{a}}\exp\left(-{s_{ab}\over s_{a}}\right).
\end{equation}
This result is consistent with random-matrix calculations 
carried out by Kogan and Kaveh \cite{Kogan95}.

The field of Eq.~(1) may be expressed as the product 
of two statistically independent random variables, 
$E$= $E^\prime$$\times$$\sqrt{s_{a}}$, where 
$E^\prime$=$E/\!\sqrt{s_{a}}$ is a Gaussian field with 
distribution $P(r^\prime,i^\prime)$=${1\over\pi}
\exp[-({r^\prime}^2$+${i^\prime}^2)]$, where 
$r^\prime$=Re[$E^\prime$] and $i^\prime$=Im$[E^\prime]$. 
This can be demonstrated by obtaining the distribution 
$P(r,i)$ of Eq.~(1) from the joint distribution 
$P(r^\prime,i^\prime;s_a)$. Since $P(r^\prime,
i^\prime;s_a)$=$P(r^{\prime},i^{\prime})$$\times$$P(s_a)$, 
the probability $\Phi_{E}(X,Y)$ that the real and imaginary 
parts of $E$ assume values less than or equal to $X$ and $Y$, 
respectively, can be expressed as $\Phi_{E}(X,Y)$= 
$\int_0^\infty\Phi_{E^\prime}({X\over\sqrt{s_a}},
{Y\over\sqrt{s_a}})P(s_{a})ds_a$. 
Its derivative is the joint probability density function, 
$P_{E}(X,Y)\equiv{d^2\Phi_{E}(X,Y)\over dX\,dY}$= 
$\int_0^\infty{1\over s_a}P_{E^\prime}({X\over\sqrt{s_a}},
{Y\over\sqrt{s_a}})P(s_a)ds_a$, which is in agreement with Eq.~(1). 

The normalized intensity $s_{ab}$ can also be 
written as a product of two statistically independent variables, 
$s_{ab}$=$s_{ab}^{\prime}$$\times$$s_{a}$, where 
$s_{ab}^{\prime}$=$|E^\prime|^2\equiv{r^\prime}^{2}$+${i^\prime}^{2}$. 
The corresponding joint distribution is $P(s_{ab}^{\prime}; 
s_{a})$= $P(s_{ab}^{\prime})$$\times$$P(s_{a})$, where 
$P(s_{ab}^{\prime})$=$\exp(-s_{ab}^{\prime})$. Since the 
$n$-th moment of the product of two statistically independent 
random variables is the product of their $n$-th moments, the 
moment relation between the field and the total transmission is 
\begin{equation}
\langle\alpha^{n}\rangle=\langle{\alpha^\prime}^{n}\rangle\langle 
s_a^{n/2}\rangle=
\left\{\begin{array}{ll}{(2k-1)!!\over 2^k}\langle s_a^{k}\rangle, & n=2k \\
                           0, & n=2k-1\, , 
        \end{array}\right.
\end{equation}
where $\alpha^\prime$=$r^\prime,i^\prime$ and $k$ is an integer, 
and the moment relation between the intensity and the total 
transmission is
\begin{equation} 
\langle s_{ab}^{n}\rangle=
\langle {s^\prime_{ab}}^{\!\!\!n}\rangle\langle 
s_{a}^{n}\rangle=n!\langle s_{a}^{n}\rangle\, .
\end{equation}
Equation~(4) is in agreement with the calculations of 
Ref.~\cite{Kogan95} and measurements of Ref.~\cite{Marin99}.

To demonstrate that $P(r,i)$ is the mixture of Gaussian 
distributions of Eq.~(1), we solve for $P(s_a)$ by utilizing 
the field average $\langle\cos(2\alpha\sqrt{z})\rangle$, 
with $z$$\ge$0. Expanding the cosine as a power series in 
$\alpha$ and then using the moment relation of Eq.~(3), we 
obtain the Laplace transform of $P(s_a)$, $F(z)$, 
$\langle\cos(2\alpha\sqrt{z})\rangle$=$\langle\exp(-s_{a}z)
\rangle$$\equiv$$F(z)$. The solid curves in Fig.~2a are plots 
of $F(z)$=$\langle\cos(2\alpha\sqrt{z})\rangle$ versus $z$ for
 measurements A, B, and C. These curves are inverted to obtain 
$P(s_a)$, by using approximate inversion of the Laplace transform 
\cite{Weeks,Steh}. The distributions $P(s_a)$ are shown by the 
solid curves in Fig.~2b. 

\begin{figure}[b!]
\includegraphics[width=\columnwidth]{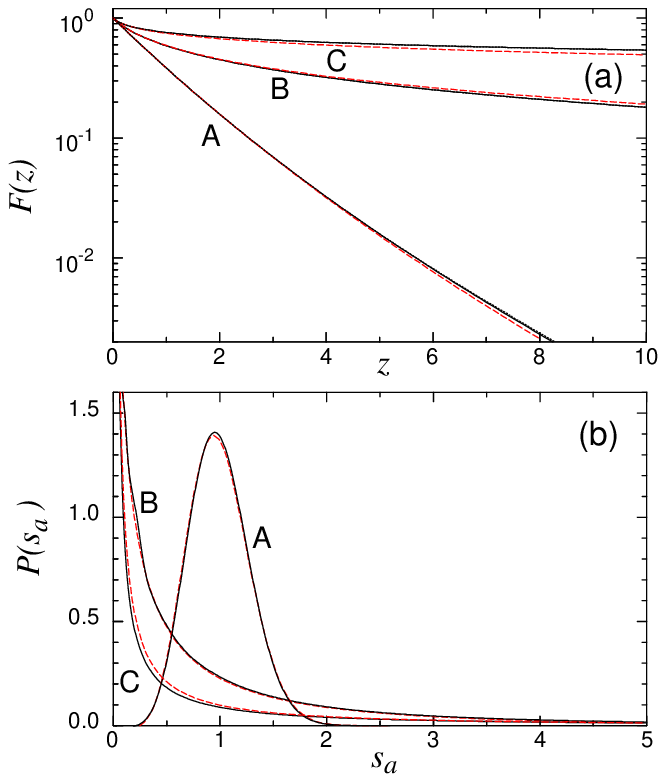}
\caption{(a) Laplace transform $F(z)$ of the total 
transmission distribution $P(s_{a})$, found as the 
field and intensity average, 
$F(z)$=$\langle\cos(2\alpha\sqrt{z})\rangle$ (solid) and 
$F(z)$=$\langle J_{0}(2\sqrt{zs_{ab}})\rangle$ (dotted), 
respectively, in measurements A, B, and C. The dashed curves 
are $F(z)$ of Eq.~(5) \cite{Rossum95,Kogan95} with $g$ replaced 
by 2/3var$(s_a)$. (b) $P(s_{a})$ found by inverting the 
corresponding $F(z)$ in (a). For measurement A, $F(z)$ is 
fit by $F(z)=\exp[Q(z)]$, where $Q(z)$ is a polynomial of 
power 3, and the resulting function is inverted using the 
Weeks method \cite{Weeks}; for measurements B and C, $F(z)$ 
is inverted using the Stehfest method \cite{Steh}.}
\end{figure}

$F(z)$ is also equal to the intensity average 
$\langle J_{0}(2\sqrt{zs_{ab}})\rangle$, where $J_{0}(x)$ 
is the Bessel function of zero order. This can be shown by 
expanding the Bessel function as a power series in $s_{ab}$ 
and then using the moment relation of Eq.~(4). 
$F(z)$=$\langle J_{0}(2\sqrt{zs_{ab}})\rangle$ for each 
measurement is plotted as the dotted curve in Fig.~2a and 
coincides with $F(z)$=$\langle\cos(2\alpha\sqrt{z})\rangle$.

$F(z)$ was previously found in diagrammatic \cite{Rossum95} 
and random-matrix-theory \cite{Kogan95} calculations in the 
diffusive limit $g$$\gg$1, in the absence of absorption, 
to be a function only of the dimensionless conductance $g$, 
\begin{equation}
F(z)=\exp[-g\ln^{2}(\sqrt{1+z/g}+\sqrt{z/g})].
\end{equation}
A nonperturbative result for $F(z)$ was obtained in 
Ref.~\cite{Beenakker96} for the case of broken time-reversal 
symmetry. Theoretical expressions for $P(s_{a})$ and $P(s_{ab})$ 
derived using Eq.~(5) were found to closely match the measured 
distributions \cite{Marin97,Marin99,Polar04}, even at the 
localization threshold, $g\sim1$, and even in the presence 
of absorption, once $g$ was replaced by 2/3var$(s_a)$. 
Var$(s_{a})$ can be found from the relation, 
var$(s_a)$=${1\over2}[\textrm{var}(s_{ab})-1]$, following 
from Eq.~(4). The plots of $F(z)$ of Eq.~(5) with $g$ 
replaced by 2/3var$(s_a)$ with the values of var$(s_a)$ 
of 0.09, 2.59 and 9.55 for measurements A, B, and C, 
respectively, are displayed as the dashed curves in Fig.~2a. 
The theoretical curves are seen to deviate from $F(z)$ found 
as the field and intensity average at large $z$. Deviations 
also appear in $P(s_{a})$ for $s_{a}<1$ (Fig.~2b); increasing 
deviations from the approximate theory of 
Ref.~\cite{Rossum95,Kogan95} for $P(s_{a})$ are observed for 
the distributions with larger values of var$(s_{a})$.

We now consider field and intensity correlation versus 
shifts in the position and polarization of the detected 
wave, $p$, and in the incident frequency or positions of 
scatterers within the medium, $q$. By virtue of the 
stationarity and statistical independence of the normalized 
field $E^\prime(p,q)$ and the total transmission $s_{a}(q)$, 
the field correlation function with shifts in $p$ and $q$, 
$\Delta{p}$ and $\Delta{q}$, 
$F_{E}(\Delta p,\Delta q)$=$\langle E(p,q)\,
E^*(p$+$\Delta p,q$+$\Delta q)\rangle$, 
can be written as the product of two correlation functions, 
\begin{equation} 
F_{E}(\Delta p,\Delta q)=F_{E^\prime}(\Delta p,\Delta q)
\times\Gamma_{\!\sqrt{s_{a}}}(\Delta q)\, ,
\end{equation}
where $F_{E^\prime}(\Delta p,\Delta q)$=$\langle E^\prime(p,q)\,
{E^\prime}^*(p$+$\Delta p,q$+$\Delta q)\rangle$ and 
$\Gamma_{\!\sqrt{s_a}}(\Delta q)$=$\langle\sqrt{s_a(q)\,
s_a(q+\Delta q)}\rangle$. Similarly, the intensity correlation 
function, $\Gamma_{s_{ab}}(\Delta p,\Delta q)$=$\langle s_{ab}(p,q)\,
s_{ab}(p$+ $\Delta p,q$+$\Delta q)\rangle$, can be written as
\begin{equation}
\Gamma_{s_{ab}}(\Delta p,\Delta q)=
\Gamma_{s_{ab}^\prime}(\Delta p,\Delta q)\times\Gamma_{s_{a}}(\Delta q)\, ,
\end{equation}
where $\Gamma_{s_{ab}^\prime}(\Delta p,\Delta q)$=$\langle 
s_{ab}^\prime(p,q)\,s_{ab}^\prime(p$+$\Delta p,q$+$\Delta q)\rangle$ 
and $\Gamma_{s_a}(\Delta q)$=$\langle s_{a}(q)\,s_{a}(q$+$\Delta q)\rangle$. 
The Siegert relation for the Gaussian random process $E^{\prime}(p,q)$ 
gives $\Gamma_{s_{ab}^\prime}(\Delta p,\Delta q)$= 1+$F^{\prime}
(\Delta p,\Delta q)$, where $F^{\prime}(\Delta p,\Delta q)$=$|F_{E^\prime}
(\Delta p,\Delta q)|^2$ \cite{Goodman}. 

For a shift in position or polarization, $\Delta p$, 
with $\Delta q$=0, $\Gamma_{\sqrt{s_a}}(\Delta q)$=$\langle 
s_{a}\rangle$=1 and $\Gamma_{s_a}(\Delta q)$=$\langle s_a^2\rangle$. 
This gives $F_E(\Delta p)$=$F_{E^\prime}(\Delta p)$ and 
$\Gamma_{s_{ab}}(\Delta p)$=$\langle s_a^2
\rangle[1$+$F^{\prime}(\Delta p)]$. 
For a shift in position, $\Delta p$=$\Delta\mathrm{r}$, 
the functional form of $F_{E^\prime}(\Delta\mathrm{r})$ 
is that predicted by coherence theory \cite{Wolf,Shapiro}, 
as found in microwave \cite{Polar04,Patrick02} and numerical 
\cite{Yamilov04} studies of field correlation. The cumulant 
intensity correlation function with displacement is given by 
$C(\Delta\mathrm{r})$=$\Gamma_{s_{ab}}(\Delta\mathrm{r})$$-$ $1$=$F^\prime(\Delta\mathrm{r})$+var$(s_a)[1$+$F^\prime(\Delta\mathrm{r})]$. 
Identifying the degree of nonlocal correlation, $\kappa$, 
as the value of $C$ when $F^\prime(\Delta\mathrm{r})$=0, 
we find $\kappa$ is precisely equal to var$(s_a)$ and thus $C(\Delta\mathrm{r})$=$F^\prime(\Delta\mathrm{r})$+$\kappa[1$+$F^\prime
(\Delta\mathrm{r})]$ \cite{Polar04,Yamilov04}. Similar results 
were found for correlation with rotation of the polarization 
of the detected field, $\Delta p$=$\Delta\theta$ \cite{Polar04}. 
For any value of var$(s_a)$, 
$F_{E^\prime}(\Delta\theta)$=$\cos(\Delta\theta)$. 
The cumulant intensity correlation function is 
$C(\Delta\theta)$=$F^\prime(\Delta\theta)$+$\kappa[1$+$F^\prime
(\Delta\theta)]$. 

When the total transmission $s_a$ varies with the incident 
frequency $\nu$, or with time $\tau$ as the internal structure 
of the sample changes, mesoscopic correlation is no longer 
represented by a single parameter. For example, the field 
and intensity correlation functions with frequency shift become $F_E(\Delta\nu)$=$F_{E^\prime}(\Delta\nu)$$\times$$\Gamma_{\!
\sqrt{s_a}}(\Delta\nu)$ 
and $\Gamma_{s_{ab}}(\Delta\nu)$=$[1$+$F^\prime
(\Delta\nu)]$$\times$$\Gamma_{s_a}(\Delta\nu)$, respectively. 
The field correlation function $F_E(\Delta\nu)$ is of 
particular interest because it is the Fourier transform 
of the time-of-flight distribution, $P(t)$, where $t$ 
is the time delay following a short pulse \cite{Drake}. 
Since $F_E(\Delta\nu)$ is the product of two functions, 
$P(t)$ is the convolution of their two Fourier transforms. 
In addition, the correlation function 
$\Gamma_{\!\sqrt{s_a}}(\Delta\nu)$ can be written as 
$\Gamma_{\!\sqrt{s_a}}(\Delta\nu)$=${\langle\sqrt{s_a}\,
\rangle}^2$+$\delta\Gamma_{\!\sqrt{s_a}}(\Delta\nu)$, 
with $\delta\Gamma_{\!\sqrt{s_a}}(\Delta\nu)$ vanishing 
at large $\Delta\nu$. We therefore expect that $P(t)$ 
is the sum of two terms. The first term is associated with 
spectral correlation of the Gaussian field $E^{\prime}$, 
and the second is due to correlation of the square root 
of the total transmission.

In geometries other than quasi-1D, such as a slab, in which 
transverse modes are not completely mixed, we may replace 
the total transmission $s_a$ with a variable $s_{a}(\textbf{r})$, 
which is the average intensity within a region on the output 
surface over which field statistics is nearly Gaussian. 
The field and intensity distributions are then still given 
by Eq.~(1) and (2), respectively, and their spatial 
correlation functions are the products of two terms. 
The first term is a narrow function of $\Delta r$, 
describing correlation within the speckle pattern, 
while the second term is a broad function, reflecting 
correlation of the more slowly varying function 
$s_{a}(\textbf{r})$ over the output surface of the sample.

In conclusion, we have found the manner in which field 
statistics arise from fluctuations and correlation of 
the spatially averaged flux in mesoscopic samples. 
We find that the probability distribution of the field in 
a given quasi-1D sample configuration and in the subset 
of configurations of flux $s_a$ is Gaussian and that the field 
distribution in the random ensemble is the mixture of 
Gaussian functions weighted by the distribution $P(s_a)$ 
[Eq.~(1)]. Thus, non-Gaussian mesoscopic statistics of 
the field arise from the breakdown of ergodicity in which 
the Gaussian field statistics for different points within 
a given configuration differ from the mesoscopic statistics 
at a fixed point in a random ensemble. When the field is 
normalized by $\sqrt{s_a}$ in each configuration, it becomes 
a Gaussian random process with position, polarization, 
frequency, and time. The field correlation with displacement and 
polarization rotation is independent of closeness to the 
localization threshold or of the degree of correlation, 
$\kappa$. In contrast, the field correlation function 
with frequency or time shift is written as a product of correlators 
of the Gaussian field and $\sqrt{s_a}$. Since the 
time-of-flight distribution for particles is the Fourier 
transform of the field correlation function with frequency 
shift, the increasing suppression of transport with time delay
as a result of the increasing impact of weak localization 
is associated with mesoscopic fluctuations.

We thank S.~Bobkov, H.~Cao, and A.~Yamilov for valuable discussions. 
This research was sponsored by the National Science Foundation 
(DMR0205186) and U.S. Army Research Office (DAAD190010362).

\end{document}